\documentclass[runningheads]{llncs}

\usepackage[T1]{fontenc}
\usepackage{graphicx}
\usepackage{amsmath,amssymb}
\usepackage{booktabs}
\usepackage{multirow}
\usepackage{array}
\usepackage{algorithm}
\usepackage{algorithmic}
\usepackage{url}

\title{Robust Residual Finite Scalar Quantization for Neural Compression}
\titlerunning{Robust Residual FSQ for Neural Compression}

\author{Xiaoxu Zhu\inst{1} \and
Dongchuan Ran\inst{2} \and
Yiming Ren\inst{3} \and
Baoxiang Li\inst{4}}
\authorrunning{X. Zhu et al.}

\institute{
XPeng Motors\\
\email{zhuxx12@xiaopeng.com}
\and
Independent Researcher\\
\email{rdc01234@163.com}
\and
Beijing Tongming Lake Information Technology Application Innovation Center\\
(TLAIC)\\
\email{renyiming@tlaic.cn}
\and
Shanghai Artificial Intelligence Laboratory\\
\email{libaoxiang@pjlab.org.cn}
}

\begin{document}
\maketitle

\begin{abstract}
Finite Scalar Quantization (FSQ) is attractive for neural compression because
it removes learned codebook maintenance and usually achieves stable code
utilization. Its direct residual extension, however, is poorly matched to
low-bitrate speech coding: after early stages explain the dominant encoder
features, later stages receive small residuals that occupy only a narrow part
of the fixed FSQ grid. We study this residual magnitude decay problem and
propose Robust Residual FSQ (RFSQ), a multi-stage FSQ framework with two
conditioning strategies. Learnable scale conditioning amplifies each residual
before quantization and maps the quantized output back to the original scale,
requiring no additional transmitted symbols. Invertible LayerNorm
conditioning further standardizes residual shape and provides an empirical
upper bound on the value of full residual normalization. On 24-kHz speech
coding at 1.8 kbps, unconditioned residual FSQ obtains 3.187 DNSMOS, scale
conditioning improves it to 3.421, and LayerNorm reaches 3.646, exceeding a
strong RVQ baseline at 3.518. A small MOS study with 15 listeners shows the
same trend. On ImageNet reconstruction, LayerNorm reduces L1 and LPIPS losses
by 9.7\% and 17.4\% relative to unconditioned residual FSQ. These results show
that residual conditioning is essential when replacing RVQ codebooks with
fixed scalar quantizers.

\end{abstract}

\section{Introduction}
\label{sec:intro}

Low-bitrate neural speech codecs are now used in speech communication,
generative speech models, and speech-to-speech interfaces. Systems such as
SoundStream and EnCodec show that an encoder, a residual vector quantizer
(RVQ), and a neural decoder can reconstruct perceptually plausible audio at
bitrates far below classical waveform codecs~\cite{soundstream,encodec}. The
quantizer is a central part of this pipeline. It determines how effectively
the continuous encoder features are converted into a discrete bitstream and
how much information remains available to the decoder.

RVQ is effective because it decomposes the quantization problem into a
sequence of residual stages. Each stage only has to explain the error left by
previous stages, so the total representation becomes progressively more
accurate as more codebooks are used. This design is especially suitable for
speech, where coarse spectral structure, speaker characteristics, prosody,
and fine waveform details naturally occupy different scales. Nevertheless,
standard RVQ relies on learned vector codebooks. In practice, these codebooks
often require commitment losses, exponential moving-average updates,
restart heuristics, and careful balancing to avoid collapse or severe
under-utilization~\cite{vqvae,robusttraining}. These training complications
are manageable, but they increase engineering complexity and can make
comparisons across codecs sensitive to implementation details.

Finite Scalar Quantization (FSQ) offers a simpler alternative. It quantizes
each feature dimension independently to a fixed set of scalar levels, thereby
avoiding a learned vector codebook altogether~\cite{fsq}. Because the
Cartesian product of the scalar levels defines the implicit codebook, FSQ can
achieve high code utilization without codebook-specific losses. This property
has made FSQ and closely related lookup-free quantizers useful in visual
tokenization, speech synthesis, automatic speech recognition, and low-bitrate
audio modeling~\cite{lfq,cosyvoice2,codecasr,scalingtransformers}. A natural
question is whether FSQ can replace VQ codebooks inside a residual
quantization chain and inherit the advantages of both approaches.

The straightforward residual extension does not work as well as this
intuition suggests. Once the first stage has captured the high-energy part of
the encoder feature, the next residual is much smaller than the original
signal. Later residuals can be smaller again. FSQ's fixed scalar grid is then
applied to signals whose dynamic range is only a fraction of the intended
range. The quantizer is not saturated; instead, most of its levels are idle.
This failure mode differs from classical clipping or range violation. The
residuals remain inside the quantizer's support, but they do not use enough of
the support to spend the nominal number of bits effectively. We refer to this
as \emph{residual magnitude decay}.

This paper proposes Robust Residual Finite Scalar Quantization (RFSQ), a
conditioned residual FSQ framework designed to counteract magnitude decay. We
study two strategies. The first is learnable scale conditioning, which
multiplies a residual by a positive learned scale before FSQ and divides the
quantized output by the same scale before updating the residual. This is the
minimal intervention needed to align residual magnitude with the scalar grid,
and its scale parameters are ordinary model weights. The second is invertible
LayerNorm conditioning, which standardizes each residual before quantization
and applies the inverse affine transform to the quantized value. This
normalizes both magnitude and distribution shape, and it is useful for
understanding the performance that full residual standardization can provide.
As discussed in Section~\ref{sec:limitations}, the LayerNorm variant also
raises a stricter deployment question because exact index-only decoding must
reproduce or transmit the normalization statistics.

Our experiments focus on the NCMMSC-relevant setting of speech and audio
coding, with image reconstruction used as a cross-domain stress test. In
24-kHz speech coding at 1.8 kbps, RFSQ with LayerNorm reaches 3.646 DNSMOS,
whereas unconditioned residual FSQ obtains 3.187 and a strong RVQ baseline
obtains 3.518. Scale conditioning improves the unconditioned system to 3.421,
recovering a large part of the degradation without adding transmitted
symbols. ImageNet reconstruction shows the same pattern: conditioned residual
FSQ consistently improves over the unconditioned residual chain, especially
on perceptual LPIPS loss.

The main contributions are as follows.
\begin{enumerate}
  \item We identify residual magnitude decay as a key obstacle when FSQ is
  used in multi-stage residual quantization.
  \item We formulate RFSQ with scale and invertible LayerNorm conditioning,
  providing a simple way to adapt each residual stage to its effective
  operating range.
  \item We evaluate RFSQ on low-bitrate speech coding with objective DNSMOS,
  a small MOS study, conditioning ablations, bit-allocation ablations, and
  stage-count ablations.
  \item We verify that the same conditioning principle improves image
  reconstruction, indicating that the problem is not specific to a single
  speech architecture.
\end{enumerate}

\section{Related Work}
\label{sec:related}

\subsection{Vector Quantization for Neural Compression}

Vector quantization has long been a basic tool for signal compression and
discrete representation learning~\cite{gray1984vector}. VQ-VAE introduced a
practical neural version in which an encoder output is replaced by the
nearest entry in a learned codebook and gradients are passed through the
quantization operation with a straight-through estimator~\cite{vqvae}. The
approach enabled discrete latent-variable models for audio, images, and
video, but it also exposed several training issues. A learned codebook can
collapse to a small subset of entries, and unused entries receive little or no
gradient signal. Later work introduced commitment terms, exponential
moving-average updates, codebook restarts, and other stabilization
mechanisms~\cite{vqvae2,robusttraining}.

For neural audio compression, residual vector quantization has become the
dominant choice. SoundStream and EnCodec employ multiple quantizers in a
residual chain, allowing each codebook to encode a progressively smaller
error signal~\cite{soundstream,encodec}. The Descript Audio Codec further
improves RVQ-based audio coding through stronger adversarial training and
multi-scale discriminator design~\cite{dac}. These systems show that residual
decomposition is highly effective for audio, but they retain the complexity
of learned vector codebooks. Our goal is not to replace all elements of these
systems; rather, it is to understand whether the codebook-free simplicity of
FSQ can be combined with the multi-stage structure that makes RVQ effective.

\subsection{Codebook-Free and Scalar Quantization}

FSQ quantizes each dimension independently to a small set of fixed scalar
levels~\cite{fsq}. Its implicit codebook is the Cartesian product of these
levels, so codebook entries are not learned parameters. Lookup-Free
Quantization (LFQ) pushes a similar idea further by using binary or ternary
per-dimension assignments and has been used successfully in visual
tokenizers~\cite{lfq}. The appeal of these approaches is practical: they
reduce the number of moving parts in the quantizer and avoid codebook collapse
by construction.

Recent speech systems have started to adopt scalar or lookup-free discrete
representations. CosyVoice 2 uses FSQ-style discrete units in speech
synthesis~\cite{cosyvoice2}; Codec-ASR studies automatic speech recognition
from codec-like discrete speech representations~\cite{codecasr}; and recent
low-bitrate speech coding work explores the scaling behavior of transformer
codecs and discrete speech tokens~\cite{scalingtransformers}. These successes
motivate a closer look at residual FSQ. The key difference in our setting is
that each later quantizer operates on a residual rather than on a fresh
encoder feature. This residual structure changes the input distribution of
each stage and creates the magnitude decay problem addressed here.

\section{Robust Residual Finite Scalar Quantization}
\label{sec:method}

\subsection{Finite Scalar Quantization}

Let $\mathbf{z}\in\mathbb{R}^{d}$ be an encoder feature vector. FSQ assigns
dimension $i$ to one of $L_i$ scalar levels. In the normalized interval
$[-1,1]$, the scalar quantizer can be written as
\begin{equation}
  \operatorname{FSQ}_i(z_i)
  =
  \operatorname{round}\left(\frac{z_i(L_i-1)}{2}\right)
  \frac{2}{L_i-1}.
  \label{eq:fsq}
\end{equation}
The implicit codebook size is $\prod_{i=1}^{d}L_i$, and the nominal rate is
$\sum_{i=1}^{d}\log_2 L_i$ bits per token. Because the scalar dimensions are
addressed independently, the model does not need to learn, update, or restart
high-dimensional codebook entries. Here $i$ indexes the scalar dimension,
$d$ is the number of scalar dimensions, $z_i$ is the $i$-th scalar input, and
$L_i$ is the number of scalar levels assigned to that dimension.

In practice, the encoder and decoder around FSQ learn to produce features
whose useful range matches the fixed grid. This calibration works well for a
single-stage quantizer, but it becomes fragile in a residual chain because
later stages see distributions that are determined by the errors of earlier
stages rather than by the original encoder feature distribution.

\subsection{Residual Magnitude Decay}

A naive residual FSQ chain applies $K$ scalar quantizers sequentially:
\begin{align}
  \mathbf{q}_1 &= \operatorname{FSQ}_1(\mathbf{z}),
  &
  \mathbf{r}_1 &= \mathbf{z}-\mathbf{q}_1,
  \label{eq:naive_first}\\
  \mathbf{q}_k &= \operatorname{FSQ}_k(\mathbf{r}_{k-1}),
  &
  \mathbf{r}_k &= \mathbf{r}_{k-1}-\mathbf{q}_k,\quad k>1.
  \label{eq:naive_later}
\end{align}
Here $K$ is the number of residual stages, $\mathbf{q}_k$ denotes the
stage-$k$ contribution in the original feature space, and $\mathbf{r}_k$
denotes the residual after subtracting the first $k$ stages.
If the first stage is effective, then $\Vert\mathbf{r}_1\Vert$ is much
smaller than $\Vert\mathbf{z}\Vert$. Empirically, we observe a decay pattern
close to
\begin{equation}
  \Vert\mathbf{r}_k\Vert \approx \rho_k\Vert\mathbf{r}_{k-1}\Vert,
  \qquad 0<\rho_k<1,
  \label{eq:decay}
\end{equation}
with later residuals frequently occupying a small central interval of the
FSQ grid. The coefficient $\rho_k$ is the empirical residual-energy ratio
between two consecutive stages. The effective number of used scalar levels can
therefore be far below $L_i$.

Consider a one-dimensional quantizer with eight levels over $[-1,1]$. The
nominal rate is three bits. If a residual lies mostly in $[-0.25,0.25]$, only
the central levels are visited regularly. The bitstream still spends three
nominal bits for that dimension, but the quantizer behaves closer to a
one-bit quantizer for the actual residual distribution. In a residual chain,
this inefficiency compounds because the later stages are precisely the ones
responsible for fine perceptual details.

This observation leads to a design criterion: each residual stage should see
an input distribution that covers the intended FSQ operating range. RFSQ
implements this criterion through conditioning transforms that are applied
before FSQ and inverted after FSQ.

\subsection{Scale Conditioning}

Scale conditioning uses a positive stage-specific scale $\alpha_k$ to
compensate for the typical reduction in residual magnitude. We distinguish
$\mathbf{u}_k$, the FSQ output in the scaled space, from $\mathbf{q}_k$, the
same contribution mapped back to the original residual space:
\begin{align}
  \tilde{\mathbf{r}}_{k-1} &= \alpha_k\mathbf{r}_{k-1}, \quad \alpha_k>0,
  \label{eq:scale_input}\\
  \mathbf{u}_k &= \operatorname{FSQ}_k(\tilde{\mathbf{r}}_{k-1}),
  \label{eq:scale_quant}\\
  \mathbf{q}_k &= \mathbf{u}_k / \alpha_k,
  \label{eq:scale_inverse}\\
  \mathbf{r}_k &= \mathbf{r}_{k-1}-\mathbf{q}_k.
  \label{eq:scale_residual}
\end{align}
The scale is parameterized to remain positive, for example through a
softplus-transformed learned parameter. The inverse in
Eq.~\eqref{eq:scale_inverse} maps the quantized value back to the original
residual scale before subtraction. Thus, the conditioning transform itself
does not add approximation error; all approximation comes from FSQ in the
conditioned space.

Scale conditioning is bitstream-friendly. The scales are model parameters
shared by the encoder and decoder, just like convolution weights. They do not
vary across samples and do not have to be transmitted for each frame. This
makes the scale variant a conservative deployment path for residual FSQ.

\subsection{Invertible LayerNorm Conditioning}

Magnitude correction alone does not address all distributional mismatch.
Later residuals can have non-zero means, unequal variances across dimensions,
and asymmetric shapes induced by earlier quantization errors. We therefore
also study an invertible LayerNorm conditioning strategy. Let $\mathcal{A}_k$
be the axes normalized by stage $k$; for sequence features this is the
stage's scalar feature axis, while for image feature maps it can be the
stage's spatial or channel-wise normalization axis. The statistics used by
stage $k$ are computed from the current residual $\mathbf{r}_{k-1}$. As in
scale conditioning, $\mathbf{u}_k$ denotes the FSQ output in the conditioned
space and $\mathbf{q}_k$ denotes the corresponding contribution after inverse
conditioning:
\begin{align}
  \boldsymbol{\mu}_k &= \operatorname{mean}_{\mathcal{A}_k}(\mathbf{r}_{k-1}),
  &
  \boldsymbol{\sigma}_k &=
  \sqrt{\operatorname{var}_{\mathcal{A}_k}(\mathbf{r}_{k-1})+\epsilon},
  \label{eq:ln_stats}\\
  \hat{\mathbf{r}}_{k-1} &=
  \operatorname{LN}_k(\mathbf{r}_{k-1})
  =
  (\mathbf{r}_{k-1}-\boldsymbol{\mu}_k)
  \oslash \boldsymbol{\sigma}_k,
  \label{eq:ln_forward}\\
  \mathbf{u}_k &= \operatorname{FSQ}_k(\hat{\mathbf{r}}_{k-1}),
  \label{eq:ln_quant}\\
  \mathbf{q}_k &=
  \operatorname{LN}_k^{-1}(\mathbf{u}_k)
  =
  \mathbf{u}_k\odot\boldsymbol{\sigma}_k+\boldsymbol{\mu}_k,
  \label{eq:ln_inverse}\\
  \mathbf{r}_k &= \mathbf{r}_{k-1}-\mathbf{q}_k.
  \label{eq:ln_residual}
\end{align}
Here $\epsilon>0$ is a numerical stabilizer. The operators $\odot$ and
$\oslash$ denote element-wise multiplication and division, with
$\boldsymbol{\mu}_k$ and $\boldsymbol{\sigma}_k$ broadcast to the residual
shape. Optional learnable affine gain and bias terms can be treated as part
of $\operatorname{LN}_k$ and inverted by $\operatorname{LN}_k^{-1}$; they are
omitted from the displayed equations to keep the notation aligned with the
residual FSQ update. This transform forces the input to each scalar quantizer
to have a more stable dynamic range and distribution shape.

LayerNorm conditioning should be interpreted carefully in codec deployment.
In the experiments, the normalization statistics are available inside the
quantized reconstruction path, so the comparison isolates the value of full
residual standardization. A standalone index-only decoder must either
reproduce these statistics from a shared state, approximate them with learned
or running statistics, or transmit compact side information. For this reason,
we report LayerNorm as the best-performing normalization strategy and discuss
its deployment caveat explicitly in Section~\ref{sec:limitations}. The scale
variant is the stricter no-side-information version.

\subsection{Stage Allocation}

Residual quantization does not require all stages to receive the same number
of bits. Early stages see higher-energy residuals and therefore benefit from
more capacity. We evaluate both uniform and non-uniform allocations. The best
speech configuration uses four stages with $(8,6,5,5)$ bits, totaling
24 bits per frame. The corresponding bitrate at 75 frames/s is
\begin{equation}
  24~\text{bits/frame}\times 75~\text{frames/s}=1800~\text{bits/s}.
  \label{eq:bitrate}
\end{equation}
This allocation is compared against a uniform four-stage allocation
$(6,6,6,6)$ and against two-stage and eight-stage alternatives.

\begin{algorithm}[t]
\caption{Robust Residual FSQ}
\label{alg:rfsq}
\begin{algorithmic}[1]
\REQUIRE Encoder feature $\mathbf{z}$; stages $K$; strategy in
\{\texttt{none}, \texttt{scale}, \texttt{layernorm}\}
\ENSURE Quantized feature $\mathbf{q}_{\mathrm{tot}}$ and stage indices
$\{\mathbf{I}_1,\ldots,\mathbf{I}_K\}$
\STATE $\mathbf{r}_0 \leftarrow \mathbf{z}$;
$\mathbf{q}_{\mathrm{tot}}\leftarrow \mathbf{0}$
\FOR{$k=1$ \textbf{to} $K$}
  \IF{strategy $=$ \texttt{scale} \textbf{and} $k>1$}
    \STATE $\mathbf{u}_k,\mathbf{I}_k \leftarrow
    \operatorname{FSQ}_k(\alpha_k\mathbf{r}_{k-1})$
    \STATE $\mathbf{q}_k \leftarrow \mathbf{u}_k/\alpha_k$
  \ELSIF{strategy $=$ \texttt{layernorm} \textbf{and} $k>1$}
    \STATE $\hat{\mathbf{r}}_{k-1} \leftarrow
    \operatorname{LN}_k(\mathbf{r}_{k-1})$
    \STATE $\mathbf{u}_k,\mathbf{I}_k \leftarrow
    \operatorname{FSQ}_k(\hat{\mathbf{r}}_{k-1})$
    \STATE $\mathbf{q}_k \leftarrow
    \operatorname{LN}_k^{-1}(\mathbf{u}_k)$
  \ELSE
    \STATE $\mathbf{q}_k,\mathbf{I}_k \leftarrow
    \operatorname{FSQ}_k(\mathbf{r}_{k-1})$
  \ENDIF
  \STATE $\mathbf{r}_k \leftarrow \mathbf{r}_{k-1}-\mathbf{q}_k$
  \STATE $\mathbf{q}_{\mathrm{tot}}\leftarrow
  \mathbf{q}_{\mathrm{tot}}+\mathbf{q}_k$
\ENDFOR
\RETURN $\mathbf{q}_{\mathrm{tot}}$,
$\{\mathbf{I}_1,\ldots,\mathbf{I}_K\}$
\end{algorithmic}
\end{algorithm}

\section{Experimental Setup}
\label{sec:setup}

\subsection{Speech Coding Task}

The main evaluation task is 24-kHz speech reconstruction at 1.8 kbps. We use
a clean subset of the Emilia multilingual speech corpus~\cite{emilia}. Audio
is downsampled to 24 kHz. The encoder-decoder follows the SEANet-style design
used by EnCodec~\cite{encodec}, with four convolutional downsampling blocks
and strides $[2,4,5,8]$. The product of the strides is 320, so the latent
sequence rate is $24000/320=75$ frames/s. Each frame is represented by a
128-dimensional encoder feature before the quantizer.

All compared systems use the same encoder-decoder architecture, the same
training objective, and the same nominal index budget of 24 bits/frame. The
loss combines waveform, spectral, adversarial, and feature-matching terms:
\begin{align}
  \mathcal{L} &=
  \lambda_{\mathrm{time}}\|\mathbf{x}-\hat{\mathbf{x}}\|_1
  +\lambda_{\mathrm{stft}}\mathcal{L}_{\mathrm{STFT}}
  +\lambda_{\mathrm{spec}}\mathcal{L}_{\mathrm{spec}} \notag\\
  &\quad
  +\lambda_{\mathrm{adv}}\mathcal{L}_{\mathrm{adv}}
  +\lambda_{\mathrm{feat}}\mathcal{L}_{\mathrm{feat}},
  \label{eq:audio_loss}
\end{align}
with $\lambda_{\mathrm{time}}=1.0$, $\lambda_{\mathrm{stft}}=1.0$,
$\lambda_{\mathrm{spec}}=0.1$, $\lambda_{\mathrm{adv}}=1.0$, and
$\lambda_{\mathrm{feat}}=2.0$. In Eq.~\eqref{eq:audio_loss},
$\mathbf{x}$ and $\hat{\mathbf{x}}$ denote the input waveform and the
reconstructed waveform, respectively; $\mathcal{L}_{\mathrm{STFT}}$ is the
multi-resolution STFT loss, $\mathcal{L}_{\mathrm{spec}}$ is the spectral
loss, $\mathcal{L}_{\mathrm{adv}}$ is the adversarial loss, and
$\mathcal{L}_{\mathrm{feat}}$ is the discriminator feature-matching loss.
Models are trained for 200k steps with batch size 32. The network has
approximately 25M parameters.

\subsection{Speech Baselines and RFSQ Variants}

Table~\ref{tab:configs} summarizes the quantizer configurations. The
baselines include VQ with exponential moving-average updates and product
quantization (VQ-EMA-4$\times$64-PQ), LFQ-24D~\cite{lfq}, single-stage FSQ
with levels $[64,64,64,64]$, and RVQ-4$\times$64. The RVQ baseline is the
most direct comparison because it uses the same residual decomposition
principle as RFSQ but with learned vector codebooks.

The RFSQ variants test three questions: whether residual conditioning is
necessary, whether scale or LayerNorm is more effective, and how the number
of stages and bit allocation affect performance. RFSQ-4S-NU-LN is the main
four-stage non-uniform system with LayerNorm conditioning and $(8,6,5,5)$
bits. RFSQ-4S-NU-Scale and RFSQ-4S-NU-No keep the same allocation but replace
the conditioning. RFSQ-4S-Uni-LN uses four uniform 6-bit stages. RFSQ-2S-NU-LN
and RFSQ-8S-Uni-LN test coarser and finer stage granularities.

\begin{table}[t]
\centering
\caption{Speech quantizer configurations under a 24 bits/frame budget.}
\label{tab:configs}
\small
\setlength{\tabcolsep}{5pt}
\begin{tabular}{llc}
\toprule
Method & Configuration & Bits/frame \\
\midrule
VQ-EMA-4$\times$64-PQ & 4 product sub-codebooks, 64 entries each & 24 \\
LFQ-24D & 24 binary/lookup-free dimensions & 24 \\
FSQ-4D-Uniform & single FSQ, levels $[64,64,64,64]$ & 24 \\
RVQ-4$\times$64 & 4 learned VQ stages, 64 entries each & 24 \\
\midrule
RFSQ-2S-NU-LN & 2 stages, non-uniform, LayerNorm & 24 \\
RFSQ-4S-NU-No & 4 stages, $(8,6,5,5)$ bits, no conditioning & 24 \\
RFSQ-4S-NU-Scale & 4 stages, $(8,6,5,5)$ bits, scale & 24 \\
RFSQ-8S-Uni-LN & 8 stages, uniform, LayerNorm & 24 \\
RFSQ-4S-Uni-LN & 4 stages, $(6,6,6,6)$ bits, LayerNorm & 24 \\
RFSQ-4S-NU-LN & 4 stages, $(8,6,5,5)$ bits, LayerNorm & 24 \\
\bottomrule
\end{tabular}
\end{table}

\subsection{Evaluation Metrics}

We use DNSMOS as the main objective speech quality metric~\cite{dnsmos_reddy}.
DNSMOS predicts subjective quality scores from a non-intrusive neural model
and is commonly used for noise suppression and speech enhancement evaluation.
Although it is not a replacement for formal listening tests, it enables
consistent large-scale comparison across quantizer variants. To complement
DNSMOS, we also report a small MOS study averaged over 15 listeners. The MOS
results should be interpreted as supporting evidence rather than as a full
MUSHRA-style subjective evaluation.

\subsection{Image Reconstruction Task}

We include ImageNet reconstruction as a cross-domain test of the residual
conditioning principle~\cite{imagenet}. Images are resized to
128$\times$128. The encoder uses two stride-2 convolutional layers with
4$\times$4 kernels, producing 32$\times$32 feature maps; the decoder mirrors
this structure with transposed convolutions. The loss is
\begin{equation}
  \mathcal{L}_{\mathrm{img}} =
  \lambda_1\|\mathbf{x}-\hat{\mathbf{x}}\|_1
  +\lambda_p\operatorname{LPIPS}(\mathbf{x},\hat{\mathbf{x}}),
  \label{eq:image_loss}
\end{equation}
where $\lambda_1=\lambda_p=1.0$ and LPIPS is computed with a learned
perceptual metric~\cite{lpips}. In Eq.~\eqref{eq:image_loss}, $\mathbf{x}$
and $\hat{\mathbf{x}}$ denote the input image and reconstructed image,
respectively, and $\lambda_1$ and $\lambda_p$ weight pixel-domain and
perceptual reconstruction terms. We compare conditioned and unconditioned RFSQ
at 22-bit and 40-bit settings.

\section{Results and Analysis}
\label{sec:results}

\subsection{Speech Coding Results}

Table~\ref{tab:speech_results} reports DNSMOS and MOS. The unconditioned
residual FSQ system performs poorly despite using the same 24 bits/frame
budget as the other systems. It reaches 3.187 DNSMOS, which is 9.4\% below
the RVQ baseline. This supports the core claim that residual decomposition
alone is insufficient when the later residuals are fed directly into fixed
FSQ grids.

Scale conditioning raises DNSMOS from 3.187 to 3.421. This is a 7.3\%
relative gain over unconditioned residual FSQ, achieved with the same index
budget and without per-frame side symbols. The result indicates that a large
part of the failure is indeed caused by magnitude mismatch. However, scale
conditioning still trails RVQ by 2.8\%, suggesting that global magnitude
alignment cannot completely correct distributional mismatch.

LayerNorm conditioning gives the best objective result. RFSQ-4S-NU-LN obtains
3.646 DNSMOS, improving over RVQ-4$\times$64 by 3.6\% and over
RFSQ-4S-NU-No by 14.4\%. The MOS results follow the same trend: the uniform
LayerNorm variant has the highest MOS at 3.83, and the non-uniform LayerNorm
variant obtains 3.80, both above RVQ's 3.81 within the resolution of this
small study. These subjective results are not large enough to establish
statistical significance, but they are consistent with the objective
evaluation.

\begin{table}[t]
\centering
\caption{Objective DNSMOS and small-scale subjective MOS evaluation for
speech coding at 1.8 kbps. MOS results are averaged from 15 listeners.}
\label{tab:speech_results}
\small
\setlength{\tabcolsep}{7pt}
\begin{tabular}{lccc}
\toprule
Method & DNSMOS & MOS & DNSMOS vs. RVQ \\
\midrule
VQ-EMA-4$\times$64-PQ & $2.687 \pm 0.468$ & 3.42 & $-23.6\%$ \\
LFQ-24D & $2.814 \pm 0.437$ & 3.52 & $-20.0\%$ \\
FSQ-4D-Uniform & $2.965 \pm 0.383$ & 3.51 & $-15.7\%$ \\
RVQ-4$\times$64 & $3.518 \pm 0.281$ & 3.81 & $0.0\%$ \\
\midrule
RFSQ-2S-NU-LN & $3.289 \pm 0.296$ & 3.62 & $-6.5\%$ \\
RFSQ-4S-NU-No & $3.187 \pm 0.319$ & 3.58 & $-9.4\%$ \\
RFSQ-4S-NU-Scale & $3.421 \pm 0.274$ & 3.70 & $-2.8\%$ \\
RFSQ-8S-Uni-LN & $3.356 \pm 0.262$ & 3.65 & $-4.6\%$ \\
RFSQ-4S-Uni-LN & $3.598 \pm 0.258$ & \textbf{3.83} & $+2.3\%$ \\
\textbf{RFSQ-4S-NU-LN} & $\mathbf{3.646 \pm 0.251}$ & 3.80 & $\mathbf{+3.6\%}$ \\
\bottomrule
\end{tabular}
\end{table}

\begin{figure}[t]
\centering
\includegraphics[width=0.85\textwidth]{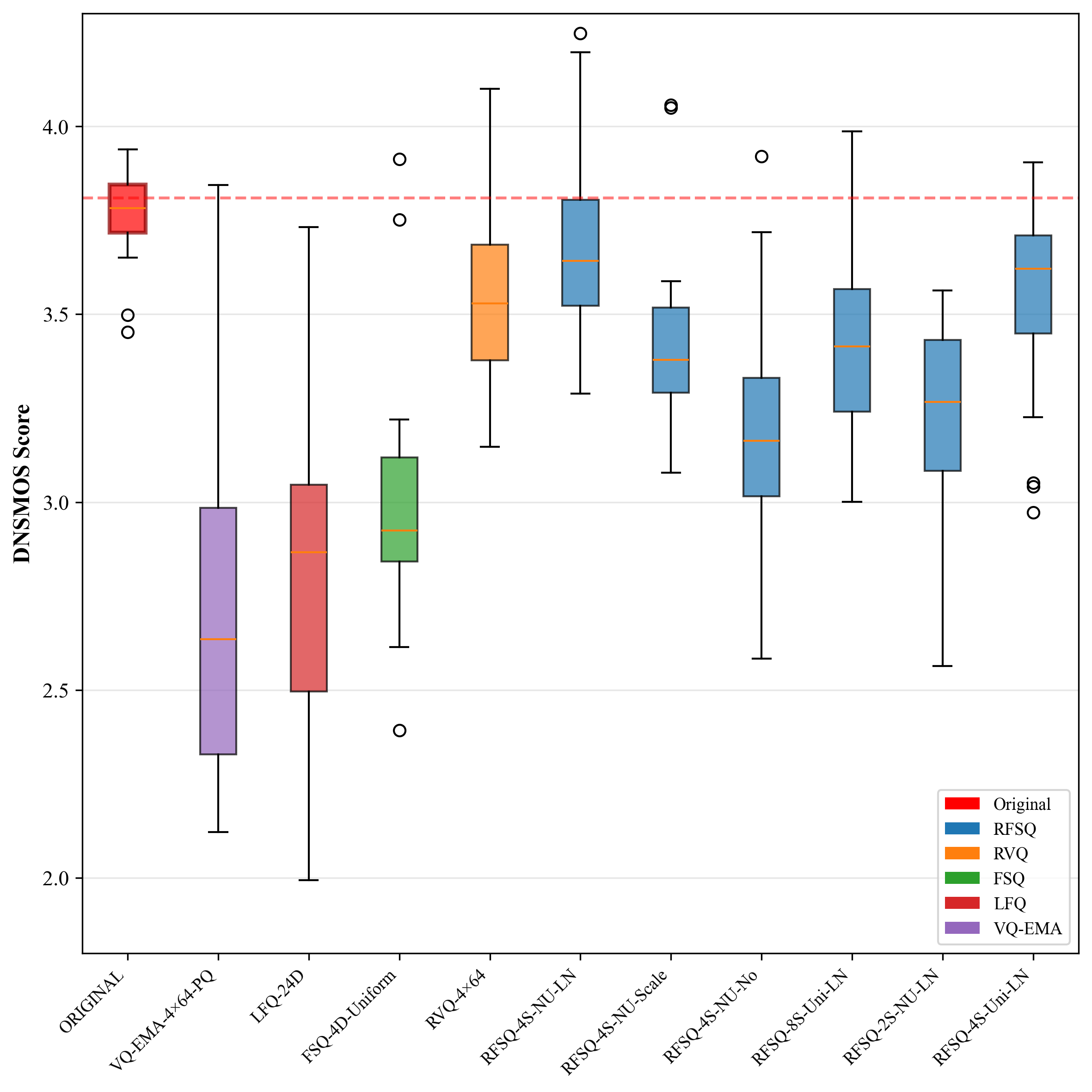}
\caption{DNSMOS score distributions for speech coding at 1.8 kbps. Box plots
show median, interquartile range, and 1.5$\times$IQR whiskers. The dashed red
line marks the original-audio DNSMOS score of 3.810.}
\label{fig:audio_dnsmos}
\end{figure}

Figure~\ref{fig:audio_dnsmos} gives the score distribution. RFSQ variants
with conditioning shift the distribution upward and reduce the lower-quality
tail relative to unconditioned residual FSQ. The lower tail matters for
speech communication because occasional severe artifacts can dominate user
experience even when the average score is acceptable. The LayerNorm variant
has both the highest mean and the smallest reported standard deviation among
the RFSQ systems in Table~\ref{tab:speech_results}.

\subsection{Ablation: Conditioning Strategy}

The conditioning ablation isolates the effect of the proposed transforms
while keeping the residual structure and bit allocation fixed. Moving from no
conditioning to scale conditioning improves DNSMOS by 0.234 absolute points.
Moving from scale conditioning to LayerNorm adds another 0.225 points. This
nearly equal split suggests that two effects are important. First, later
residuals must be amplified to use the FSQ grid. Second, after their
magnitude is corrected, their shape still differs from the distribution for
which fixed scalar levels are most effective.

The ablation also clarifies how to interpret RFSQ. The scale variant is the
strict no-side-information design and is already much stronger than naive
residual FSQ. The LayerNorm variant is the most informative design for
studying the value of full residual normalization. It demonstrates that if a
codec can provide stable normalization statistics to the decoder, residual
FSQ can surpass the RVQ baseline in this setup.

\subsection{Ablation: Stage Count and Bit Allocation}

The stage-count results show that four stages are preferable under the fixed
24 bits/frame budget. The two-stage LayerNorm variant scores 3.289, indicating
that too few stages leave too much structure in the residual. The eight-stage
variant scores 3.356, worse than the four-stage LayerNorm variants. With too
many stages, each stage receives very few bits and the accumulated
quantization error offsets the benefit of finer residual decomposition.

Bit allocation also matters. The non-uniform LayerNorm variant
RFSQ-4S-NU-LN scores 3.646, compared with 3.598 for the uniform
RFSQ-4S-Uni-LN system. Both use 24 bits/frame and four stages, so the
improvement comes from assigning more bits to earlier, higher-energy
residuals. This mirrors a classical transform-coding principle: components
with greater expected energy or perceptual importance should receive more
bits. In RFSQ, the stages themselves form such an ordered set of components.

\subsection{Image Reconstruction Results}

Table~\ref{tab:image_results} reports ImageNet reconstruction results. The
conditioning trend is consistent with the speech task. At 40 bits, the
unconditioned four-stage RFSQ obtains 0.113 L1 and 0.121 LPIPS. Scale
conditioning improves these values to 0.103 and 0.101. LayerNorm gives the
best L1 and LPIPS, 0.102 and 0.100. Relative to the unconditioned 40-bit
system, LayerNorm reduces L1 by 9.7\% and LPIPS by 17.4\%.

At 22 bits, the gains are smaller in absolute PSNR but remain visible in
LPIPS. RFSQ-2$\times$2048-LN obtains 0.148 LPIPS, compared with 0.159 for
the unconditioned variant. This suggests that residual conditioning is
especially useful for perceptually important details that occupy less energy
than coarse image structure. This is analogous to the speech setting, where
late residual stages influence fine acoustic texture and artifact suppression.

\begin{table}[t]
\centering
\caption{Image reconstruction on ImageNet 128$\times$128.}
\label{tab:image_results}
\small
\setlength{\tabcolsep}{6pt}
\begin{tabular}{lcccc}
\toprule
Method & Bits & L1 $\downarrow$ & LPIPS $\downarrow$ & PSNR $\uparrow$ \\
\midrule
RFSQ-2$\times$2048-None & 22.0 & 0.130 & 0.159 & 21.1 \\
RFSQ-2$\times$2048-Scale & 22.0 & 0.122 & 0.152 & 21.5 \\
RFSQ-2$\times$2048-LN & 22.0 & 0.124 & 0.148 & 21.3 \\
\midrule
RFSQ-4$\times$1024-None & 40.0 & 0.113 & 0.121 & 22.2 \\
RFSQ-4$\times$1024-Scale & 40.0 & 0.103 & 0.101 & 22.9 \\
\textbf{RFSQ-4$\times$1024-LN} & \textbf{40.0} & \textbf{0.102} & \textbf{0.100} & \textbf{22.9} \\
\bottomrule
\end{tabular}
\end{table}

\begin{figure}[t]
\centering
\includegraphics[width=0.9\textwidth]{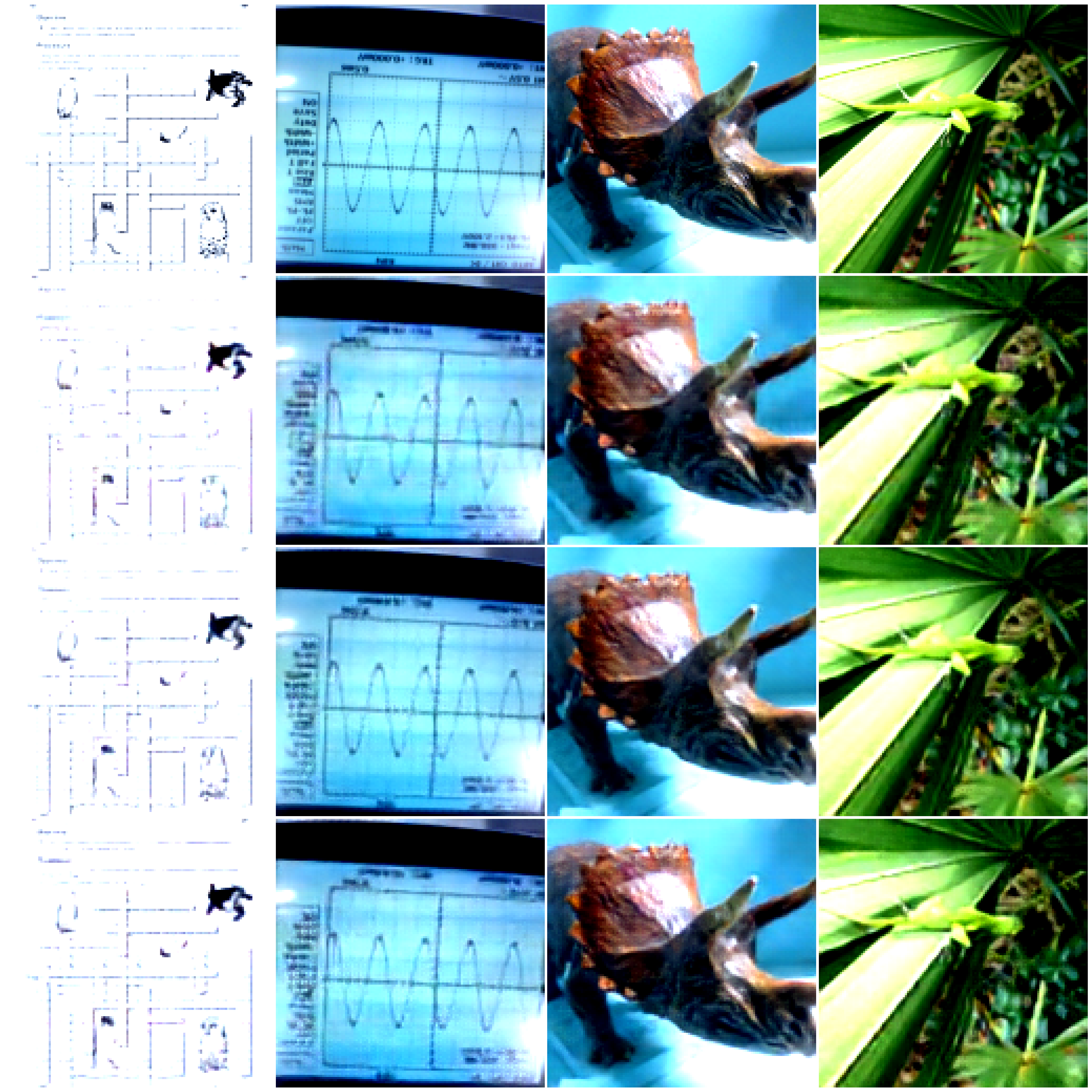}
\caption{Visual comparison on ImageNet. Rows from top to bottom: original,
RFSQ-2$\times$2048-LN at 22 bits, RFSQ-4$\times$1024-LN at 40 bits, and
RFSQ-4$\times$1024-None at 40 bits.}
\label{fig:visual}
\end{figure}

Figure~\ref{fig:visual} shows qualitative examples. The conditioned 40-bit
model preserves sharper edges and more stable texture than the unconditioned
40-bit model. The 22-bit result captures the main structure but loses more
fine detail. These visual observations are consistent with the LPIPS
improvements in Table~\ref{tab:image_results}.

\section{Discussion and Limitations}
\label{sec:limitations}

\subsection{What the Results Show}

The experiments support three conclusions. First, naive residual FSQ is not a
drop-in replacement for RVQ. Without conditioning, the later FSQ stages use
their nominal capacity inefficiently. Second, magnitude correction is a
strong baseline. The scale variant substantially improves speech quality
while keeping the bitstream interface as simple as FSQ. Third, full residual
standardization provides additional gains. The LayerNorm variant improves
both speech and image metrics, indicating that residual distribution shape
matters beyond scalar magnitude.

These conclusions are useful for neural speech codec design. If deployment
simplicity is the priority, scale-conditioned RFSQ provides a conservative
route. If the codec can support additional normalization information or a
learned approximation to the residual statistics, LayerNorm-style
conditioning indicates further headroom.

\subsection{Bitstream Accounting}

The bitrate reported for all systems is the nominal index payload:
24 bits/frame at 75 frames/s for the speech task. This is the standard first
comparison point for quantizer architectures, but it is not a complete
entropy-coded codec specification. In particular, scale conditioning requires
only shared model parameters and therefore does not change the payload. Exact
LayerNorm conditioning, as evaluated here, depends on the normalization
statistics used in the inverse transform. A deployed index-only decoder must
handle these statistics explicitly.

There are several possible deployment paths. One is to replace per-sample
statistics with learned stage-wise statistics, sacrificing some of the
adaptivity observed in the LayerNorm experiment. Another is to transmit
coarsely quantized statistics at a lower temporal rate. A third is to design a
causal prediction mechanism in which the decoder estimates normalization
statistics from already reconstructed states. The present paper does not
resolve this engineering tradeoff; it separates the no-side-information scale
result from the stronger normalization result.

\subsection{Evaluation Scope}

The subjective study is intentionally small and should not be interpreted as
a full listening test. A larger MUSHRA or ITU-T P.808-style evaluation would
be needed to quantify perceptual differences with confidence. In addition,
our RVQ baseline is strong and controlled, but it does not include every
recent training improvement from systems such as DAC~\cite{dac}. Future work
should combine RFSQ with stronger adversarial training, entropy coding, and
variable-rate operation.

\section{Conclusion}
\label{sec:conclusion}

We presented RFSQ, a conditioned residual extension of finite scalar
quantization for neural compression. The central observation is that naive
residual FSQ suffers from residual magnitude decay: later stages receive
signals that are too small and too distributionally shifted to use the fixed
FSQ grid efficiently. Scale conditioning directly corrects magnitude mismatch
and improves speech DNSMOS from 3.187 to 3.421 without adding transmitted
symbols. Invertible LayerNorm further standardizes residual shape and reaches
3.646 DNSMOS, surpassing the RVQ baseline in the controlled 1.8-kbps speech
setting. ImageNet reconstruction confirms that the same conditioning idea
reduces both distortion and perceptual loss outside speech. These results
show that residual conditioning is the key ingredient for making scalar
quantizers competitive in multi-stage neural compression. Future work should
focus on bitstream-compatible LayerNorm approximations, entropy-coded
variable-rate RFSQ, and larger subjective evaluation for speech coding.

\bibliographystyle{splncs04}
\bibliography{references}

\end{document}